\documentclass[aps,preprint,amsmath,amssymb,amsfonts,showpacs]{revtex4}
\usepackage{epsfig}
\usepackage{graphicx}
\usepackage{subfigure}
\usepackage{dcolumn}
\usepackage{bm}

\newcommand{\del}{\partial}

\renewcommand{\Re}{\operatorname{Re}}
\renewcommand{\Im}{\operatorname{Im}}

\newcommand{\calL}{\mathcal{L}}

\begin{document}

\title{The imaginary part of the gravity action and black hole entropy}

\author{Yasha Neiman}
\email{yashula@gmail.com}
\affiliation{Institute for Gravitation \& the Cosmos and Physics Department, Penn State, University Park, PA 16802, USA}

\date{\today}

\begin{abstract}
As observed recently in arXiv:1212.2922, the action of General Relativity (GR) in finite spacetime regions has an imaginary part that resembles the Bekenstein entropy. In this paper, we expand on that argument, with attention to different causal types of boundaries. This property of the GR action may open a new approach to the puzzles of gravitational entropy. In particular, the imaginary action can be evaluated for non-stationary finite regions, where the notion of entropy is not fully understood. As a first step in constructing the precise relationship between the imaginary action and entropy, we focus on stationary black hole spacetimes. There, we identify a class of bounded regions for which the action's imaginary part precisely equals the black hole entropy. As a powerful test on the validity of the approach, we also calculate the imaginary action for Lovelock gravity. The result is related to the corresponding entropy formula in the same way as in GR. 
\end{abstract}


\maketitle
\tableofcontents
\newpage

\section{Introduction} \label{sec:intro}

\subsection{Black hole entropy and stationarity}

Black hole entropy \cite{Bekenstein:1972tm,Bekenstein:1973ur,Bekenstein:1980jp,Hawking:1974rv} provides a valuable clue in the quest for a quantum theory of gravity. Our understanding of it can be evaluated along two parameters. First, it is instructive to separate special properties of the GR action from the general properties of diff-invariant field theories. In particular, one wishes to compare GR to higher-derivative theories of the metric. Second, some aspects of the entropy formula are understood for general solutions, while others are restricted to quasi-stationary setups. 

For example, in GR, the Bekenstein-Hawking entropy formula $\sigma = A/4G$ is expected to apply for quite general surfaces in general spacetimes (at least as a bound, and with various subtleties, reviewed e.g. in \cite{Bousso:2002ju}). This is supported by the fact that the area increase theorem, and with it the Second Law of black hole thermodynamics, holds even for highly non-stationary processes. Now, in higher-derivative gravity, the entropy formula must be modified, as prescribed by Wald \cite{Wald:1993nt,Jacobson:1993vj}. For general theories, Wald's procedure is sharply defined only in stationary spacetimes (however, see the proposal \cite{Iyer:1994ys} for a non-stationary formula). Also, a Second Law has only been established for quasi-stationary processes \cite{Jacobson:1995uq,Kolekar:2012tq}. It is possible that both problems may be solved at once: a correct formula for non-stationary entropy may be identified by requiring an increase law. This has been achieved for $f(R)$ theories, i.e. for Lagrangians polynomial in the Ricci scalar \cite{Jacobson:1995uq}.

In addition to the laws of black hole thermodynamics, it's important to consider more direct derivations of the entropy. Such derivations convince us that the thermodynamic interpretation is more than an analogy. They also provide us with the constant coefficient in the entropy formula. These derivations must involve at least some element of quantum mechanics, to produce the necessary $\hbar$ factor (which we neglected above, working in QFT units). In this category, we have Hawking's calculation of the black-body radiation \cite{Hawking:1974rv}, the analysis of Euclidean black holes \cite{Gibbons:1976ue}, as well as 't Hooft's S-matrix approach \cite{'tHooft:1999bw}. Also notable is the local approach in \cite{Bianchi:2012br}, based on entanglement entropy. These arguments are all restricted to stationary backgrounds. 

We see that much of our understanding of black hole entropy relies on stationarity, more so in higher-derivative theories. The scale at which stationarity is typically required is the time scale $1/\kappa$ of the inverse surface gravity (or inverse temperature). Is this a sensible situation? On one hand, it stands to reason that a thermodynamic quantity is sharply defined only for stationary systems. On the other hand, there seems to be a mismatch of the relevant scales. $1/\kappa$ is a macroscopic length, while the entropy supposedly arises from fluctuations at the Planck scale (and perhaps at other characteristic scales in a higher-derivative Lagrangian). To put this differently, both the Bekenstein and Wald formulas for the entropy give a local surface density that is independent of the surface gravity scale. It is therefore strange that $\kappa$ should be relevant to any restrictions on the entropy formula. In particular, in GR, one would imagine that \emph{any} surface with a smooth, classical geometry is sufficiently stationary and homogeneous, as far as Planck-scale fluctuations are concerned. Therefore, one should always have an entropy formula that is simply additive over such surfaces.

We conclude that it would be desirable to derive the gravitational entropy, including its $\hbar$-dependent proportionality constant, without making use of stationarity. In \cite{Neiman:2012fx}, a candidate for such a derivation was presented. There, it was pointed out that the on-shell GR action in a null-bounded region, specifically the boundary term, has an area-proportional imaginary part. This imaginary part, arising from an analytical continuation of boost angles, has the same form as the Bekenstein entropy formula. No assumption of stationarity was required for this result. The result also holds in the presence of arbitrary minimally-coupled matter, and therefore shares the ``universal'' nature of black hole entropy. On the other hand, the connection to actual black hole solutions remained unclear.

\subsection{Outline of the paper}

In the first part of this paper, we review and elaborate on the argument in \cite{Neiman:2012fx}, extending it to non-null closed boundaries in Lorentzian spacetime. In particular, we find the imaginary part of the GR action for smooth (or mixed signature) boundaries, for purely-spacelike boundaries and for purely-timelike ones. We also present a more flexible alternative approach to the analytical continuation involved. The general result for the action's imaginary part takes the form:
\begin{align}
 \Im S = \frac{1}{4}\sum_{\mathrm{flips}}\sigma_{\mathrm{flip}} \ . \label{eq:result}
\end{align}
Here, the sum is over the ``signature-flip surfaces'' where the boundary normal goes from timelike to spacelike or vice versa. These are always present for a closed boundary, though they may be hidden inside ``corners'' - surfaces where smooth patches of the boundary intersect. $\sigma_{\mathrm{flip}}$ is the entropy functional $A/4G$ evaluated on each flip surface. 

The result \eqref{eq:result} is ``empirical'', in the sense that it simply comes out from a careful evaluation of the action's boundary term. As a first step towards clarifying the precise relationship between \eqref{eq:result} and the concept of black hole entropy, we identify a particular relation that applies to stationary black holes. Specifically, for any such black hole, we indicate a class of purely-spacelike, purely-timelike and purely-null closed boundaries, such that $\Im S$ for the corresponding region evaluates precisely to the black hole entropy $\sigma$. A factor of $\hbar$ arises implicitly when the action is treated as dimensionless. 

As discussed in \cite{Neiman:2012fx}, our results complement the analysis of Euclidean black holes \cite{Gibbons:1976ue}. There, one also encounters an imaginary on-shell action in GR. Roughly, this leads to an exponential damping of transition amplitudes $e^{iS}\sim e^{-\Im S}$, which can be ascribed to a multiplicity of states. More precisely, the imaginary action in \cite{Gibbons:1976ue} is interpreted as the logarithm $\ln Z$ of a partition function. From the latter, one extracts the entropy, after correcting for terms related to conserved charges. These terms are e.g. $E/T$ for energy and $J\Omega/T$ for angular momentum ($E,T,J,\Omega$ are respectively the energy, temperature, angular momentum and angular velocity). For a Euclidean black hole, the correction terms can be viewed as contributions from the boundary at infinity, while the entropy itself is a contribution from the horizon \cite{Banados:1993qp,Carlip:1993sa}. 

The derivation here and in \cite{Neiman:2012fx} has a number of advantages over the Euclidean one. First, it traces the imaginary action directly to properties of Lorentzian geometry. The analytical continuation we employ is more subtle, with no Wick rotation of the metric. As a result, stationarity is not required: one can calculate the imaginary action for arbitrary solutions. Also, the calculation works with finite regions, with no reference to the asymptotic boundary. This is more physical, especially in a positive-$\Lambda$ cosmology. Finally, we recover the entropy formula \emph{directly} from the imaginary action, with no correction terms such as $E/T$ or $J\Omega/T$. This can be traced both to the use of finite regions (there are no ``terms at infinity'' to speak of) and to the absence of a stationarity requirement (without it, thermodynamic potentials like $T$ and $\Omega$ are not well-defined).

On the other hand, our result is incomplete. Apart from the considerations above, there is no precise understanding as to why the imaginary action should be an entropy. There is only the fact that the two quantities agree for stationary black holes. Thus, what we have is not quite a non-stationary calculation of entropy, but a non-stationary calculation of \emph{something} - the imaginary action - that agrees with the entropy in the stationary case. At the end of the paper, we speculate on the possibility that in general, the imaginary action \emph{supplants} the concept of entropy.  

In the second part of the paper, we generalize the GR calculation to higher-derivative theories. Since the action's imaginary part arises from the boundary term, we restrict to theories with a standard boundary-data formulation. For gravity without matter, the only such theories are of the Lanczos-Lovelock type \cite{Lovelock:1971yv}. Thus, we set out to find the imaginary part of the action in Lovelock gravity, using the boundary term found in \cite{Myers:1987yn}. As with GR, the result takes the form \eqref{eq:result}, where $\sigma$ is now the appropriate entropy functional for Lovelock gravity \cite{Jacobson:1993xs}. The result $\Im S = \sigma$ for a special class of regions in a stationary black hole spacetime is also recovered. 

At certain points in the higher-derivative calculation, we assume that the relevant spacetime regions are large with respect to the higher-order Lovelock couplings. The GR term then dominates, allowing us to evade the open questions regarding the positivity and increase of the entropy functional. This assumption does \emph{not} entail stationarity at the surface gravity scale. Also, the calculation remains exact (up to quantum corrections).

The paper is organized as follows. In sections \ref{sec:GR:smooth}-\ref{sec:GR:null}, we elaborate on the treatment of the GR action from \cite{Neiman:2012fx}. In section \ref{sec:GR:black_hole}, we recover the entropy of a stationary black hole from the action in suitable regions. In section \ref{sec:compare}, we consider alternative recipes for evaluating the action, while addressing some objections that our procedure may raise. In section \ref{sec:Lovelock}, we generalize these results to Lovelock gravity. Section \ref{sec:discuss} is devoted to discussion and outlook.

In the formulas and figures below, the spacetime metric $g_{\mu\nu}$ has mostly-plus signature. The spacetime dimension is $d\geq 2$. We use indices $(\mu,\nu,\dots)$ for spacetime coordinates, $(a,b,\dots)$ for coordinates on a codimension-1 hypersurface, and $(i,j,\dots)$ for coordinates on a codimension-2 surface. The sign convention for the Riemann tensor is $R^\mu{}_{\nu\rho\sigma}V^\nu = [\nabla_\rho,\nabla_\sigma]V^\mu$.

\section{The imaginary part of the GR action} \label{sec:GR}

In this section, we consider General Relativity with no higher-derivative terms. We allow for an arbitrary cosmological constant $\Lambda$ and minimally-coupled matter with Lagrangian $\mathcal{L}_M$. In this setup, black hole entropy is expected to be given by the Bekenstein-Hawking formula $\sigma = A/4G$.
 
\subsection{Smooth boundaries} \label{sec:GR:smooth}

In this subsection, we derive the imaginary part of the GR action in a region $\Omega$ with a smooth boundary. The discussion here follows and expands on that in \cite{Neiman:2012fx}. 

The complete action in the spacetime region $\Omega$ reads:
\begin{align}
 S = \frac{1}{16\pi G}\int_\Omega \sqrt{-g}\,(R + \Lambda + 16\pi G\mathcal{L}_M)\, d^dx 
  + \frac{1}{8\pi G}\int_{\del\Omega} \sqrt{\frac{-h}{n\cdot n}}\,K\, d^{d-1}x \ . \label{eq:GR_action}
\end{align}
The second integral is the Gibbons-Hawking boundary term \cite{York:1972sj,Gibbons:1976ue}, which will be our object of interest. $h$ is the determinant of the boundary metric. $K$ is the trace of the extrinsic curvature tensor $K_a^b = \nabla_a n^b$. The normal $n_\mu$ is chosen to be an outgoing covector, i.e. to have a positive scalar product with outgoing vectors. This makes the \emph{vector} $n^\mu$ outgoing, ingoing or tangent, for timelike, spacelike and null boundaries, respectively. See figure \ref{fig:angles_smooth_boundary}. The inclusion of $n\cdot n = n_\mu n^\mu$ in the denominator makes eq. \eqref{eq:GR_action} valid for normals of arbitrary length and signature. In particular, $n^\mu$ need not be a unit vector. This will facilitate the discussion of the null limit.  

\begin{figure}%
\centering%
\includegraphics[scale=0.75]{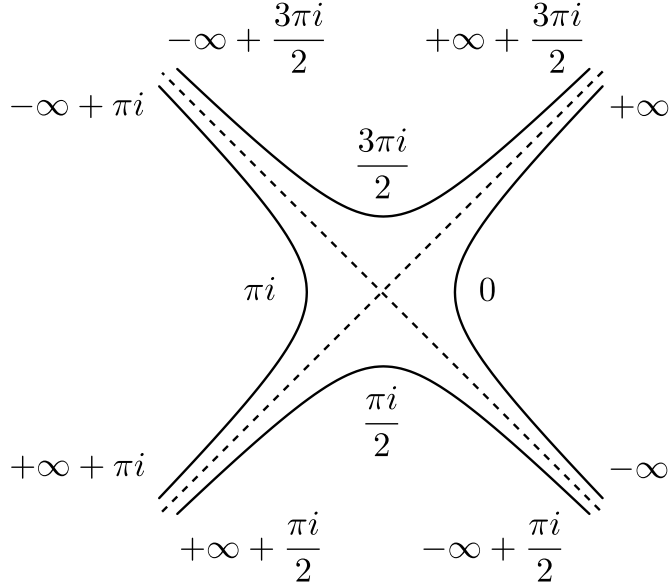} \\
\caption{An assignment of boost angles $\eta$ in the Lorentzian plane, according to eq. \eqref{eq:eta}. This is one of two complex-conjugate choices, distinguished by the sign convention \eqref{eq:sign}. The horizontal and vertical axes describe the spacelike and timelike components of a vector $n^\mu$. The angles are defined up to integer multiples of $2\pi i$.}
\label{fig:angles_plane} 
\end{figure}

\begin{figure}%
\centering%
\includegraphics[scale=0.75]{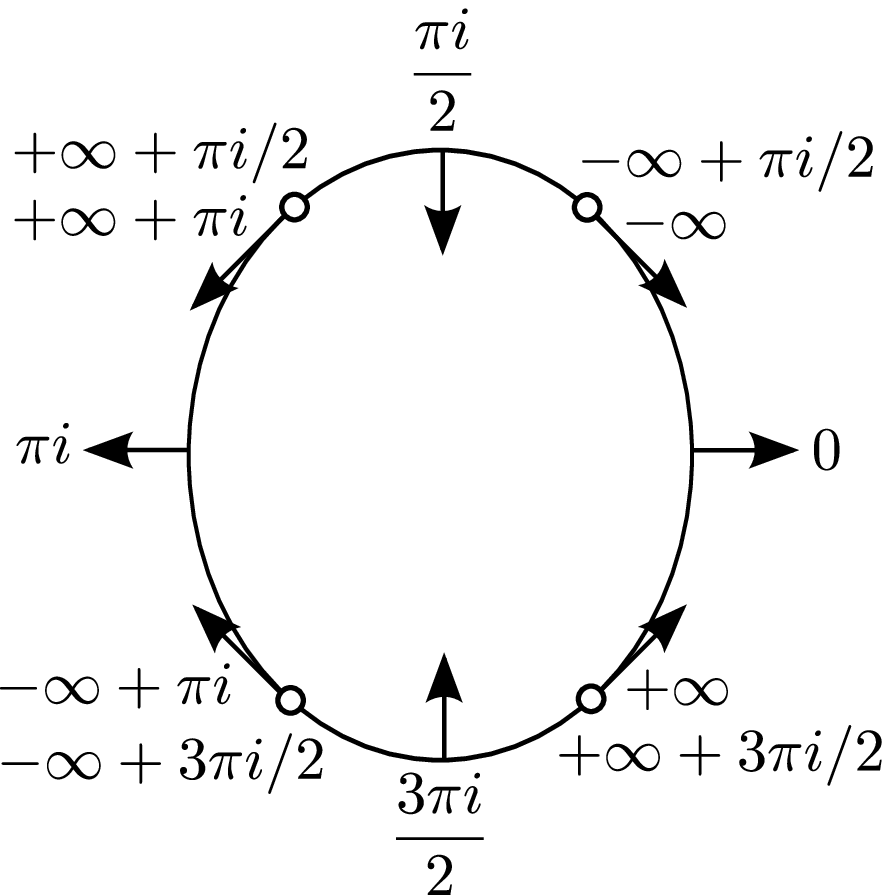} \\
\caption{The boost angles $\eta$ from figure \ref{fig:angles_plane}, as applied to the normal vectors of a smooth closed boundary in 1+1d spacetime. The arrows indicate the normal direction. The normal's sign is chosen so that it has a positive scalar product with outgoing vectors. Empty circles denote ``signature flips'', where the normal becomes momentarily null. Contributions to the Gibbons-Hawking boundary term can be read off by traveling counter-clockwise and picking up angle differences, multiplied by $1/(8\pi G)$. The angles are defined up to integer multiples of $2\pi i - \theta$, where $\theta$ is a real finite deficit angle arising from spacetime curvature.}
\label{fig:angles_smooth_boundary} 
\end{figure}

We wish to show that the boundary term in \eqref{eq:GR_action} has an imaginary part. We begin by recalling its geometric structure in flat 2d spacetime. There, the boundary $\del\Omega$ is just a 1d curve. The Gibbons-Hawking integral simply adds up the infinitesimal rotation angles of the normal $n^\mu$ as we travel along the boundary:
\begin{align}
 \int \sqrt{\frac{-h}{n\cdot n}}\,K\, dx = \int d\alpha \ . \label{eq:integral_angles}
\end{align}
In Lorentzian spacetime, this expression must be handled with care. The ``angles'' $d\alpha$ are actually boost parameters, and these can diverge. To correctly evaluate the integral over a closed boundary, one needs a consistent assignment of boost angles throughout the Lorentzian plane. Thus, we want to assign an angle $\eta$ to each vector $n^\mu = (n^0,n^1)$ in $\mathbb{R}^{1,1}$. This is non-trivial, since the angle covers the entire range from $-\infty$ to $+\infty$ within a single quadrant, e.g. $n^1 > \left|n^0\right|$. We must analytically continue $\eta$ beyond this quadrant, which unavoidably takes it into the complex plane. This can be done \cite{SorkinThesis} by analytically continuing the trigonometric formulas:
\begin{align}
  \cosh\eta = \frac{n^1}{\sqrt{n\cdot n}}; \quad \sinh\eta = \frac{n^0}{\sqrt{n\cdot n}} \ , \label{eq:eta}
\end{align}
where $n\cdot n = (n^1)^2 - (n^0)^2$, and we define:
\begin{align}
 \sqrt{n\cdot n} \equiv i\sqrt{\left|n\cdot n\right|} \quad \mbox{for} \quad n\cdot n < 0 \ . \label{eq:sign}
\end{align}
As will become clear below, the complex-conjugate choice $\sqrt{n\cdot n} \equiv -i\sqrt{\left|n\cdot n\right|}$ is ruled out by the requirement that quantum transition amplitudes $e^{iS}$ do not explode exponentially. The angle assignment \eqref{eq:eta} is illustrated in figure \ref{fig:angles_plane}. In particular, the angle $\eta$ picks up an imaginary contribution $-\pi i/2$ whenever we cross a null line counter-clockwise into a neighboring quadrant. The total angle for a complete counter-clockwise circuit is $-2\pi i$. Our conventions here differ from \cite{SorkinThesis} by an overall factor of $i$. In particular, we treat continuous boost angles within a quadrant as real. 

Let us now apply this angle assignment to the normal $n^\mu$ on our closed boundary $\del\Omega$. The boundary must be composed of both spacelike and timelike regions. See figure \ref{fig:angles_smooth_boundary}. The arrows indicate the direction of $n^\mu$ at various points. The complex numbers indicate the normal's angle in the Lorentzian plane.   

The integral \eqref{eq:integral_angles} may now be evaluated by summing up the angle differences as one travels counter-clockwise along the boundary (note that this makes $n^\mu$ rotate \emph{clockwise} - a peculiarity of Lorentzian spacetime). One can verify that this recipe produces the correct signs, by comparing with an explicit computation of $K$ for a purely spacelike or timelike patch of the boundary. Multiplying by the $1/(8\pi G)$ coefficient in \eqref{eq:GR_action}, we get the corresponding contributions to the action. 
  
In particular, figure \ref{fig:angles_smooth_boundary} shows that each time the boundary flips its signature, i.e. momentarily becomes null, the Gibbons-Hawking integral picks up an imaginary contribution of $\pi i/2$. This results in a contribution of $i/(16G)$ to the action. This behavior is the same as if the boundary turned a sharp corner by an angle $\alpha = \pi i/2$. We see that while the signature flips may be smooth in the topological sense, the boundary's angle there has an imaginary discontinuity.

This conclusion carries through beyond flat 2d spacetime. In curved 2d spacetime, the change in the angle of $n^\mu$ upon a complete circuit of the boundary is no longer $2\pi i$, but has an additional real contribution due to the Riemann curvature. Thus, one cannot consistently assign to each point a real part $\Re\eta$ as was done in figure \ref{fig:angles_smooth_boundary}. However, the angle's imaginary part behaves the same as in the flat case, picking up contributions of $\pi i/2$ at each signature flip. The divergences in the angle's real part are also unambiguous.

In $d>2$ dimensions, the Gibbons-Hawking integral no longer has the simple interpretation \eqref{eq:integral_angles}. However, when the boundary turns a sharp corner by an angle $\alpha$, the situation is \emph{effectively} two-dimensional. Such a corner is a codimension-2 surface in spacetime, with some area $A$. The extrinsic curvature component $K_\bot^\bot$ in perpendicular to the corner surface, i.e. the component ``along the bend'', has a delta-function singularity. Its integral over a small neighborhood $\delta\Sigma$ of the corner reads \cite{Hartle:1981cf,Farhi:1989yr}:  
\begin{align}
 \int_{\delta\Sigma}\sqrt{\frac{-h}{n\cdot n}}\,K_\bot^\bot\, d^{d-1}x = \alpha A \ . \label{eq:K_corner}
\end{align}
The $\alpha$ factor arises from the integral along the bend, as in \eqref{eq:integral_angles}, while the $A$ factor arises from the integral over the corner surface. The resulting contribution to the action \eqref{eq:GR_action} reads:
\begin{align}
 S_{corner} = \frac{\alpha A}{8\pi G} \ . \label{eq:S_corner}
\end{align}

Signature flips in $d>2$ also occur at codimension-2 surfaces. There should be at least two such surfaces: one at the edge of the boundary's ``initial'' spacelike patch, and one at the edge of the ``final'' spacelike patch. Each of these corresponds to a pair of points in figure \ref{fig:angles_smooth_boundary}. Let us denote the areas of these two surfaces by $A_{flip,i}$ and $A_{flip,f}$, respectively. If the flip surfaces are composed of a greater number of connected components, as in figure \ref{fig:angles_smooth_boundary}, then we interpret $A_{flip,i}$ and $A_{flip,f}$ as the \emph{total} areas of the initial and final flip surfaces. 

As we've seen in the 2d discussion, the normal's angle at the signature flips (as measured in the 1+1d plane transverse to the flip surface) has an imaginary discontinuity $\Im\alpha = \pi/2$. Thus, the flip surfaces behave as corners, and eqs. \eqref{eq:K_corner}-\eqref{eq:S_corner} apply. We conclude that the boundary term, and with it the entire action \eqref{eq:GR_action}, has the following imaginary part:
\begin{align}
 \Im S = \frac{A_{flip,i} + A_{flip,f}}{16G} \ . \label{eq:ImS_smooth}
\end{align}
This comes together with possible divergent contributions to $\Re S$, since the real part of the angle diverges at the flip surfaces. In 1+1d, where the ``areas'' are all equal to $1$, the divergences cancel out. The boundary term in that case equals $i/4G$, plus a finite real piece due to the deficit angle (which cancels against the bulk Einstein-Hilbert term).

The result \eqref{eq:ImS_smooth}, together with the convention $e^{iS}$ for quantum amplitudes, justifies our sign choice \eqref{eq:sign}. The complex-conjugate choice would have resulted in negative $\Im S$, and thus exponentially large absolute values for the amplitudes $\left|e^{iS}\right| = e^{-\Im S}$ (note that in the semiclassical regime, $\left|\Im S\right|\gg 1$). 

\subsection{Interlude: an alternative derivation} \label{sec:GR:contour}

Before proceeding, we present an additional derivation of the formula $\Im\alpha = \pi/2$ for the boundary's bending angle at a signature flip. This will provide a second perspective on the analytical continuation involved. Also, the technique presented here will prove useful in the higher-derivative calculation in section \ref{sec:Lovelock}.

\begin{figure}%
\centering%
\includegraphics[scale=0.75]{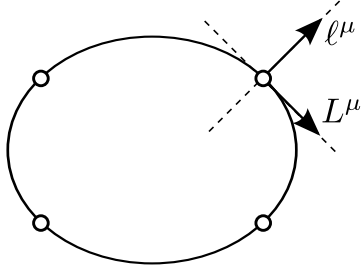} \\
\caption{A smooth closed boundary in spacetime, with flip surfaces indicated by empty circles. The dashed lines represent the two lightsheets passing through one of the flip surfaces. One lightsheet is tangent to the boundary and shares its normal vector $n^\mu \equiv L^\mu$. The other lightsheet is transverse, and has a different null normal $\ell^\mu$. The extents of the two null vectors are arbitrary.}
\label{fig:tangent_transverse} 
\end{figure}

Consider a point on one of the flip surfaces. As the flip surface is crossed, the boundary normal $n^\mu$ rotates in the orthogonal 1+1d plane. In fact, since $n^\mu$ becomes null during the signature flip, we must speak of rotation together with rescaling. The 1+1d plane is spanned by two null vectors $(L^\mu,\ell^\mu)$, each with an arbitrary extent. One of the null vectors, say $L^\mu$, is tangent (and normal) to our boundary. Thus, we may identify $L^\mu$ with the value of $n^\mu$ at the flip surface. The other null vector, $\ell^\mu$, is transverse to the boundary. See figure \ref{fig:tangent_transverse}.

In the infinitesimal neighborhood of the flip surface, we may express $n^\mu$ as a linear combination of $L^\mu$ and $\ell^\mu$:
\begin{align}
 n^\mu = \lambda(L^\mu + z\ell^\mu); \quad n\cdot n = 2\lambda^2 z (L\cdot\ell) \ , \label{eq:n_L_ell}
\end{align} 
where $\lambda\approx 1$ and $z\approx 0$. We wish to find the boost angle traced out by the vector \eqref{eq:n_L_ell} as it crosses through the null direction of $L^\mu$. In such a process, $z$ crosses through zero. Consider, then, an infinitesimal change in $n^\mu$:
\begin{align}
 dn^\mu = d\lambda L^\mu + (zd\lambda + \lambda dz)\ell^\mu; \quad dn\cdot n = \lambda(2z d\lambda + \lambda dz)(L\cdot\ell) \ .
\end{align}
The part of $dn^\mu$ orthogonal to $n^\mu$ reads:
\begin{align}
 &dn_\bot^\mu = dn^\mu - \frac{dn\cdot n}{n\cdot n}n^\mu = \frac{\lambda dz}{2}\left(\ell^\mu - \frac{1}{z}L^\mu\right); \\
 &dn_\bot\cdot dn_\bot = -\frac{\lambda^2 dz^2}{2z}(L\cdot\ell) \ . \label{eq:n_bot_squared} 
\end{align}
The square of the infinitesimal boost angle is then given by:
\begin{align}
 d\alpha^2 = -\frac{dn_\bot\cdot dn_\bot}{n\cdot n} = \frac{dz^2}{4z^2} \ . 
\end{align}
The minus sign in the second expression is related to the Lorentzian signature. It ensures a positive $d\alpha^2$ for boosts that do not involve signature flips. Note that the overall scaling $\lambda$ dropped out, as expected. For $d\alpha$, we get:
\begin{align}
 d\alpha = \pm \frac{dz}{2z} \ . \label{eq:d_alpha}
\end{align}
The sign in front of $dz/2z$ is related to the orientation of the pair $(L^\mu,\ell^\mu)$. For analyticity, it should be the same for both signs of $z$. This is consistent with the angle assignment in figure \ref{fig:angles_plane}. 

We can now find the bending angle at the signature flip as the integral of $d\alpha$ as $z$ crosses through zero:
\begin{align}
 \alpha = \pm\int_{z_-}^{z_+}\frac{dz}{2z} \ . \label{eq:alpha_integral}
\end{align}
Here, $z_-$ and $z_+$ are small real numbers, with $z_-$ negative and $z_+$ positive. To resolve the divergence at $z=0$, we deform the integration contour into the complex $z$ plane. The contour can bypass the $z=0$ pole either from above or from below, giving two complex-conjugate results. This is the same ambiguity as the one in \eqref{eq:eta}, which was resolved by \eqref{eq:sign}. The imaginary part can be found from the difference between the two contours, i.e. from a closed-contour integral around the pole:
\begin{align}
 \Im\alpha = \pm\frac{1}{2i}\oint\frac{dz}{2z} = \pm\frac{\pi}{2} \ , \label{eq:closed_contour}
\end{align}
where the two complex-conjugate options translate into the closed contour's orientation. Making the choice that leads to a positive $\Im S$, we arrive again at the result $\Im\alpha = \pi/2$. The real part of the angle \eqref{eq:alpha_integral} depends on the choice of endpoints $z_\pm$. In terms of figure \ref{fig:angles_plane}, this is just the arbitrary choice of where in the timelike and spacelike quadrants the signature flip ``begins'' and ``ends''.

\subsection{Topological corners, purely-spacelike and purely-timelike boundaries} \label{sec:GR:spacelike_timelike}

So far, our discussion revolved around smooth boundaries, where the signature-flip surfaces act as effective corners. Also of interest are boundaries with actual topological corners, i.e. surfaces where the boundary is non-differentiable. In particular, such corners are necessary if we wish to draw a closed boundary of fixed causal type. Any boundary with corners can be viewed as a limiting case of a smooth boundary. This involves ``resolving'' the corners into fast-bending smooth curves. 

\begin{figure}%
\centering%
\includegraphics[scale=0.75]{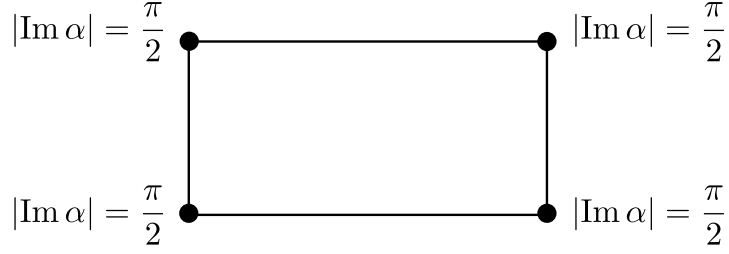} \\
\caption{A boundary of mixed signature with the flip surfaces ``hidden'' in topological corners (designated by full circles). The imaginary part of the corner angles is $\pi/2$. The angles' real part can be arbitrary. For the particular boundary depicted here, the spacelike and timelike patches are orthogonal to each other. In this case, the angles' real part is zero.}
\label{fig:mixed_boundary} 
\end{figure}

The simplest example is a boundary composed of spacelike and timelike patches, just like in figure \ref{fig:angles_smooth_boundary}, but with corners in place of the smooth flip surfaces. Such a boundary is depicted in figure \ref{fig:mixed_boundary}. The imaginary part of the GR action for the corresponding region is still given by eq. \eqref{eq:ImS_smooth}, where the flip surfaces are now identified with the corners. We note that one can reverse the limiting procedure and view the smooth boundary of figure \ref{fig:angles_smooth_boundary} as a limiting case of mixed-signature boundaries with corners. Indeed, by boosting the local frame near a corner in figure \ref{fig:mixed_boundary}, one can make the corner appear arbitrarily blunt. In this way, the boundary can be made to approach a smooth one while retaining the nonzero corner angle, and in particular its imaginary part $\pi/2$. This point of view will be important for the Lovelock-gravity calculation in section \ref{sec:Lovelock}.   

\begin{figure}%
\centering%
\includegraphics[scale=0.75]{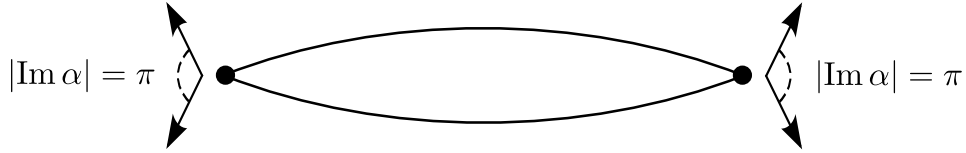} \\
\caption{A purely spacelike closed boundary, composed of two intersecting hypersurfaces. The full circles denote the intersection surface. The arrows indicate the two boundary normals at each intersection point (due to the normals' timelike signature, the future-pointing ``outgoing'' normal belongs to the initial hypersurface, and vice versa). A continuous boost between these two normals involves two signature flips. As a result, the corner angle has an imaginary part equal to $\pi$.}
\label{fig:spacelike_boundary} 
\end{figure}

As another example of a boundary with corners, figure \ref{fig:spacelike_boundary} depicts a closed spacelike boundary. It consists of two smooth spacelike hypersurfaces - one ``initial'' and one ``final''. The corner is their intersection surface. In 1+1d, this corner consists of two points. In higher dimensions, it can be a connected surface. Let us denote the corner's area by $A_0$. 
A spacelike boundary of this type can be obtained by starting from a mixed-signature boundary, as in figure \ref{fig:angles_smooth_boundary} or \ref{fig:mixed_boundary}, and compressing the whole region outside of two spacelike ``caps'' into a surface - the corner surface. In particular, both the initial and final signature-flip surfaces get compressed into the corner surface, so that $A_{flip,i} = A_{flip,f} = A_0$. As a result, the area $A_0$ is counted twice in eq. \eqref{eq:ImS_smooth}. Equivalently, as one travels around the corner in figure \ref{fig:spacelike_boundary}, the normal crosses \emph{two} quadrant borders in the Lorentzian plane, changing e.g. from future-pointing timelike to past-pointing timelike. This gives the boost angle an imaginary part $\Im\alpha = \pi$, twice larger than for a single flip. Thus, the action's imaginary part for the spacelike boundary reads:
\begin{align}
 \Im S = \frac{A_0}{8G} \ . \label{eq:ImS_spacelike}
\end{align}

Similarly, we can consider a purely timelike boundary. Its shape can be obtained by turning figure \ref{fig:spacelike_boundary} on its side. In this case, there are always two distinct corner surfaces - one initial and one final. Let us denote their areas by $A_i$ and $A_f$, respectively. Again, this boundary can be constructed from a smooth one by compressing all but a timelike region. In this process, the flip surfaces, both initial and final, get ``folded in two''. For instance, in figure \ref{fig:angles_smooth_boundary}, the initial and final pairs of flip points are each compressed into a single point. The resulting relation between the flip-surface areas of the smooth boundary and the corner areas of the timelike one is $A_{flip,i} = 2A_i$ and $A_{flip,f} = 2A_f$. Equivalently, the corners of the timelike boundary have angles with imaginary part $\Im\alpha = \pi$, just like in the purely spacelike case. We arrive at the following expression for $\Im S$:
\begin{align}
 \Im S = \frac{A_i + A_f}{8G} \ . \label{eq:ImS_timelike} 
\end{align}

The results \eqref{eq:ImS_smooth} and \eqref{eq:ImS_spacelike}-\eqref{eq:ImS_timelike} strongly resemble the Bekenstein-Hawking entropy formula, especially since they do not depend on the matter content. In section \ref{sec:GR:black_hole}, we will show how eqs. \eqref{eq:ImS_spacelike}-\eqref{eq:ImS_timelike} can be related to the entropy of a stationary black hole. We will see that the apparent factor-of-2 discrepancy between \eqref{eq:ImS_spacelike}-\eqref{eq:ImS_timelike} and the Bekenstein-Hawking formula is resolved through the presence of two intersecting horizons. Before turning to this construction, we pause to discuss the evaluation of $\Im S$ for purely null boundaries.

\subsection{Null boundaries} \label{sec:GR:null}

In this subsection, we examine the imaginary part of the on-shell GR action in a region whose boundary is everywhere null. We closely follow the argument in \cite{Neiman:2012fx}. 

Null boundaries arise naturally in the context of gravitational entropy. Null hypersurfaces have area rather than volume as a multilinear form, making them a plausible stage for area-proportional degrees of freedom. They are also crucial for generalizations of the Bekenstein bound \cite{Bousso:1999xy}. Additional arguments for null boundaries as the correct setting for quantum gravity may be found in \cite{Neiman:2012fx}. 

\subsubsection{The structure of a null boundary}

\begin{figure}%
\centering%
\includegraphics[scale=0.75]{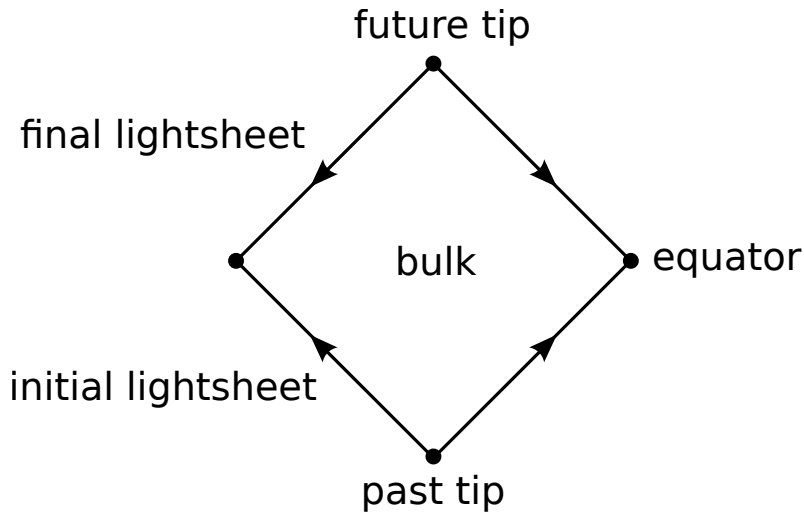} \\
\caption{Some terminology for null-bounded regions. The lines represent codimension-1 lightsheets; the dots represent codimension-2 spacelike surfaces. The arrows indicate the ``outgoing'' null normal that corresponds to the outgoing covector, for a mostly-plus metric signature. In $d>2$ dimensions with trivial topology (but not in the black-hole setup in section \ref{sec:GR:black_hole}), the left and right sides of the figure are understood to be connected.}
\label{fig:null_boundary} 
\end{figure}

A null boundary is constructed as follows (see figure \ref{fig:null_boundary} for a quick dictionary). At first, consider a spacetime metric not too different from Minkowski space. A closed null boundary is then uniquely determined by a closed codimension-2 spacelike surface. We refer to this surface as the boundary's ``equator''. The equator has precisely two lightrays orthogonal to it at each point. These two rays generate the lightcone in the 1+1d transverse timelike plane. At each point of the equator, we choose the future-pointing half of one ray and the past-pointing half of the other. Specifically, we choose the ``ingoing'' half-rays, i.e. those along which the area decreases, and which eventually intersect. Thus, we limit ourselves to ``normal'' equators in the sense of \cite{Bousso:1999xy}, and do not consider trapped or anti-trapped ones. We terminate each ray upon its first intersection with another ray. The null boundary is then given by the union of these rays. It consists of two smooth null hypersurfaces, or ``lightsheets''. Generically, the lightray intersections away from the equator constitute a pair of codimension-2 spacelike surfaces - one for the past-ingoing rays, and one for the future-ingoing ones. We refer to these surfaces as the past and future ``tips'', respectively. The initial (final) half of the boundary stretches from the equator to the past (future) tip. If the metric is a solution of GR with suitable energy conditions, then the boundary area decreases monotonously from the equator towards the tips \cite{Bousso:1999xy}. The past and future tips are then anti-trapped and trapped surfaces, respectively. In the degenerate case where the tips are single points, the two half-boundaries become lightcones. The full boundary is then a causal diamond \cite{Bousso:2000nf}.

With sufficient curvature or a non-trivial topology, existence and uniqueness issues arise. For instance, in a de-Sitter universe, equators above a certain size do not generate closed boundaries, because the lightrays never intersect. Another possibility is an equator composed of two connected components. In this case, the rays from each connected component to each tip generate a smooth lightsheet. The closed null boundary then consists of four separate lightsheets. This is always the case in 1+1d, as can be seen in figure \ref{fig:null_boundary}. In section \ref{sec:GR:black_hole}, we will encounter disjoint equators in the analysis of stationary black holes. 

We see that in terms of corner surfaces, a null boundary combines the features of a purely-spacelike and a purely-timelike one. The tips are analogous to the corners of a timelike boundary, while the equator is analogous to the corner of a spacelike boundary.

Finally, we recall the peculiar nature of normals to a null boundary. A normal vector to a lightsheet is also tangent to it. It points along the lightrays generating the lightsheet. As explained in section \ref{sec:GR:smooth}, it is still possible to define an ``outgoing'' normal, in the sense that its scalar product with outgoing vectors is positive. The direction of these normals is illustrated in figure \ref{fig:null_boundary}. They are future-pointing on initial lightsheets and past-pointing on final ones. Thus, the ``outgoing'' normal points away from tips and towards equators. 

\subsubsection{The imaginary action in a null-bounded region}

Let us come back to our object of interest, the imaginary part of the action \eqref{eq:GR_action}. This will again arise from the Gibbons-Hawking boundary term. In the null case, the evaluation of the boundary term is more subtle. A null normal $n^\mu$ of finite extent yields a finite extrinsic curvature tensor $K_a^b = \nabla_a n^b$, but has a vanishing square $n\cdot n$. On the other hand, the intrinsic metric's determinant $h$ also vanishes. As explained in \cite{Neiman:2012fx}, away from corners and flip surfaces, the Gibbons-Hawking integrand comes out finite and well-defined. We will not pursue this here, since we are only interested in the integral's imaginary part. The latter can be read off from the smooth-boundary result, up to a subtlety which is quite separate from the one mentioned above. 

\begin{figure}%
\centering%
\includegraphics[scale=0.75]{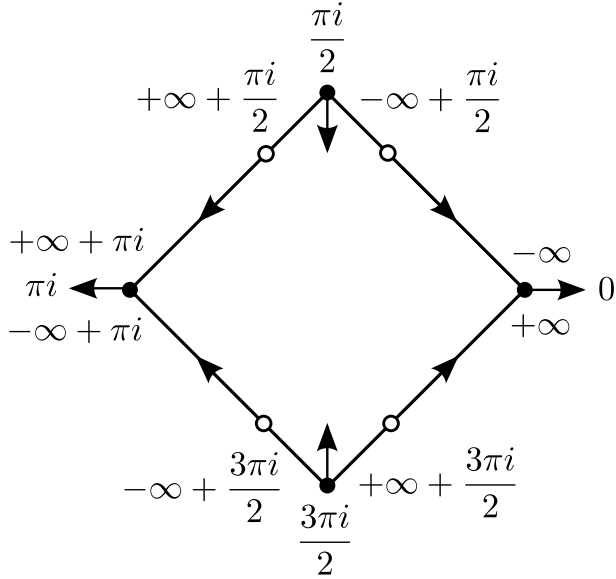} \\
\caption{The boost angles from figure \ref{fig:angles_plane}, as applied to the null boundary normals from figure \ref{fig:null_boundary}. For clarity, intermediate spacelike and timelike normals are included at the equator and tips, respectively. Empty circles denote the ``signature flip'' surfaces, where the boost angle crosses into a neighboring quadrant. When area is not constant along the lightsheets, the path integral favors placing these surfaces infinitesimally close to the tips.}
\label{fig:angles_null_boundary} 
\end{figure}

Figure \ref{fig:angles_null_boundary} is the null analogue of figure \ref{fig:angles_smooth_boundary}. We again start in 1+1d for simplicity. A null boundary in 1+1d is always a causal diamond. We again indicate the ``boost angle'' of the normal at various points, with the same qualifications as before regarding curvature-induced ambiguities. We include spacelike and timelike normals at the equator and tips, in order to clarify the behavior of the angle's real part. These non-null normals appear when we ``resolve'' the corners into fast-bending smooth curves. 

In the non-null case, we've seen that the boundary term gets imaginary ``corner contributions'' from surfaces where the normal is momentarily null. In figure \ref{fig:angles_null_boundary}, on the other hand, the normal is null throughout each lightsheet. In other words, it sits right on the border between quadrants in the Lorentzian plane (figure \ref{fig:angles_plane}). At some point, at or between the lightsheet intersections, that border must be crossed. The normal's signature then flips, and the boost angle picks up its imaginary $\pi i/2$ contribution. In $d>2$ dimensions, this must happen separately along each ray, so that overall we get a signature-flip \emph{surface} on every lightsheet. In figure \ref{fig:angles_null_boundary}, these are represented by empty circles. We denote the areas of the flip surfaces on the initial and final lightsheets by $A_{flip,i}$ and $A_{flip,f}$, respectively. As in section \ref{sec:GR:smooth}, the boundary term picks up an imaginary part at these surfaces. Its value is again given by eq. \eqref{eq:ImS_smooth}. The problem specific to the null case is that we do not know the flip surfaces' locations and areas.

To fix the locations of the flip surfaces, we imagine a path integral over the different possibilities. The imaginary part of the action suppresses the amplitudes $e^{iS}$ exponentially, as $e^{-\Im S}$. This selects the smallest possible imaginary action, just like in Euclidean path integrals. Thus, we should give $A_{flip,i}$ and $A_{flip,f}$ the smallest possible values. Now, it is known \cite{Bousso:1999xy} that on-shell, the boundary area decreases monotonously from the equator towards the tips (and will continue decreasing until a caustic is reached). Thus, the flip surfaces must be chosen at the tips. This fixes $A_{flip,i} = 2A_i$ and $A_{flip,f} = 2A_f$, where $A_i$ and $A_f$ are the initial and final tip areas. The factor of 2 arises from the two rays intersecting at each tip point, each of which carries a signature flip. Alternatively, the boundary area infinitesimally close to the tip is twice the tip area. In this respect, a null boundary behaves like a timelike one. With $A_{flip,i}$ and $A_{flip,f}$ determined in this way, $\Im S$ is given by the same formula as eq. \eqref{eq:ImS_timelike}:
\begin{align}
 \Im S = \frac{A_i + A_f}{8G} \ . \label{eq:ImS_null} 
\end{align}

If one regards \eqref{eq:ImS_null} as an entropy formula, one notices a peculiar feature. It is usually assumed that the entropy bound is concerned with the \emph{largest} available area, in particular in the null context \cite{Bousso:1999xy}. In contrast, eq. \eqref{eq:ImS_null} refers to the \emph{smallest} available area on each lightsheet. We find the resulting picture more compelling: the smallest area defines a bottleneck on the available information. For instance, a causal diamond with vanishing tip areas (in $d>2$) is associated with a pointlike observer \cite{Bousso:2000nf}. Now, a truly pointlike observer cannot carry any information at all. In particular, he cannot design an experiment in his causal diamond, or collect the results of measurements. From this perspective, it stands to reason that the entropy bound for a causal diamond should be zero.

In a limiting case, a lightsheet's area may remain constant as one travels from the equator towards the tips. We will encounter this situation in section \ref{sec:GR:black_hole}. For such a lightsheet, the path-integral argument cannot fix the location of the flip surface. This does not affect the result \eqref{eq:ImS_null}, since the flip surface's area is now independent of its location. Thus, for the purpose of finding $\Im S$, we may still think of the flip surface as located at the tip.

\subsection{Relation with stationary black holes} \label{sec:GR:black_hole}

We will now make direct contact between the general formulas \eqref{eq:ImS_spacelike}-\eqref{eq:ImS_null} and the Bekenstein entropy in stationary black hole spacetimes. Specifically, for any stationary black hole with horizon area $A$, we will indicate three families of spacetime regions for which $\Im S = A/4G$. This subsection should be viewed as merely a first step in establishing the relation between $\Im S$ and black hole entropy. Closer contact with existing derivations of the entropy is the subject of current work.  

\begin{figure}%
\centering%
\includegraphics[scale=0.75]{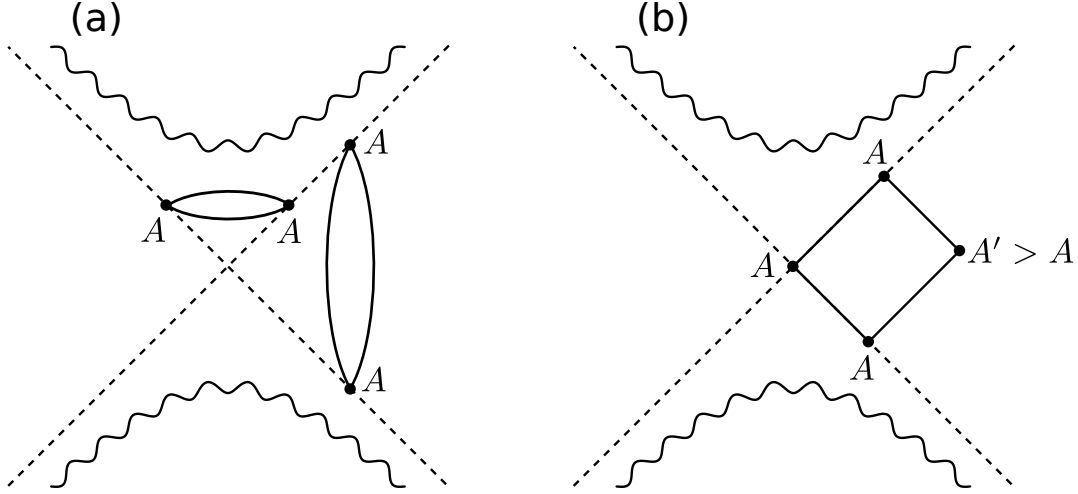} \\
\caption{A Kruskal diagram for a stationary Schwarzschild metric. The dashed lines are the two Killing horizons. Superposed on the diagram are examples of closed boundaries - spacelike and timelike in (a), null in (b) - for which the imaginary part of the GR action reproduces the black hole entropy $A/4G$. Solid lines represent hypersurfaces; solid circles represent corners. The left-hand corner of the boundary in (b) is the bifurcation surface. The top and bottom corners in (b), as well as all corners in (a), are arbitrary slices of the Killing horizons. The corner surfaces are all isometric, except for the right-hand corner in (b).}
\label{fig:kruskal} 
\end{figure}

For concreteness, consider a Schwarzschild black hole. Figure \ref{fig:kruskal} depicts the maximal extension of this spacetime using Kruskal coordinates. Of the four quadrants in the Kruskal diagram, we will be interested in the right-hand one and the top one. These are the ``physical'' outside universe and the black hole's interior, respectively. The unphysical second universe and the white hole's interior are irrelevant, as are the singularities. However, we make crucial use of the bifurcation surface and the existence of a second event horizon. Thus, the construction described here does not apply to realistic black holes formed by gravitational collapse.

In figure \ref{fig:kruskal}(a), we depict two closed boundaries - a purely spacelike boundary inside the black hole and a purely timelike boundary in the external spacetime. The corner surfaces, denoted by solid circles, are arbitrary slices of the two event horizons. For simplicity, figure \ref{fig:kruskal}(a) is drawn as if each slice is at constant Kruskal coordinates $(u,v)$. This need not be the case. Since the horizons are Killing, their slices are all isometric. In particular, they all have the same area $A$, which is just the horizon area. The solid lines in figure \ref{fig:kruskal}(a) are arbitrary smooth hypersurfaces (of fixed causal type) that stretch between the horizon slices. Thus, the spacelike boundary has a corner surface with total area $A_0 = 2A$, composed of two connected components. Substituting in eq. \eqref{eq:ImS_spacelike}, we find that the GR action for the corresponding region satisfies $\Im S = A/4G$. The same result is obtained for the timelike boundary, using eq. \eqref{eq:ImS_timelike} with the initial and final corner areas $A_i = A_f = A$.

Figure \ref{fig:kruskal}(b) depicts a similarly constructed closed \emph{null} boundary. This leads to the same result as the spacelike and timelike families, but is arguably more ``tailored'' to the black hole spacetime. The boundary's equator is composed of two connected components, one of which is the bifurcation surface. From the bifurcation surface, we draw two lightsheets along the two Killing horizons. We terminate each lightsheet at an arbitrary horizon slice. These slices form the boundary's tips. From each tip, we then draw the unique second lightsheet containing it, and follow these lightsheets until they intersect. The intersection surface forms the second component of the equator. The horizon slices, including the bifurcation surface, are all isometric with area $A$. The equator's second component - the right-hand corner in figure \ref{fig:kruskal}(b) - is a surface with some other area $A'$, enveloping the black hole from the outside. From the marginally-(anti)-trapped nature of the two horizons, we know that the area of the two right-hand lightsheets increases as one begins to follow them from the tips. We \emph{assume} that it continues increasing until the lightsheets intersect. One way to satisfy this assumption is to place the tip surfaces at constant Kruskal coordinates $(u,v)$. The assumption implies in particular that the area of the right-hand corner satisfies $A' > A$. A null boundary constructed in this way falls into the category considered in section \ref{sec:GR:null}. On the right-hand lightsheets, we will have flip surfaces at the tips, each with area $A$. On the left-hand lightsheets, where the area is constant, the flip surfaces' location is not determined. In any case, the GR action for the null-bounded region satisfies eq. \eqref{eq:ImS_null}, with tip areas $A_i = A_f = A$. We again arrive at the Bekenstein formula $\Im S = A/4G$. In particular, the result does not depend on the larger area $A'$. 

The above arguments clearly extend to stationary black holes other than Schwarzschild. The only requirements are a ``physical'' outside region, an interior region and a bifurcate Killing horizon. Moreover, to construct the timelike and null boundaries, we need only the outside region and the two horizons bounding it.

We see that the factor-of-2 discrepancy between eqs. \eqref{eq:ImS_spacelike}-\eqref{eq:ImS_null} and the Bekenstein-Hawking formula is resolved by having \emph{two} horizon slices as the relevant corner surfaces. This should raise a suspicion that the correct numerical factor is accidental, and could have come out otherwise. However, in a stationary black hole spacetime, it's unclear how one might choose a region such that $\Im S$ would still be proportional to $A/G$, but with a different numerical factor. In particular, if one wishes to avoid the second horizon, one must still place a second corner surface somewhere. This would make $\Im S$ depend on the new corner's area, rather than just on the horizon area $A$. Incidentally, it is also unclear how to draw a \emph{smooth} closed boundary in figure \ref{fig:kruskal} such that the imaginary action \eqref{eq:ImS_smooth} would depend solely on the horizon area.

We note that one could also draw a null boundary in the \emph{top} quadrant of figure \ref{fig:kruskal}, such that the two equator components, rather than the two tips, are horizon slices with area $A$. The initial tip would then be the bifurcation surface, and the final tip would be an arbitrary interior surface with area $A'' < A$. For such a boundary, eq. \eqref{eq:ImS_null} would give $\Im S = (A + A'')/8G < A/4G$, failing to reproduce the black hole entropy. 

\section{Comparison with other evaluations of the action} \label{sec:compare}

In section \ref{sec:GR}, we derived the imaginary part of the GR action by a ``brute force'' evaluation of the boundary integral in \eqref{eq:GR_action}. Some objections may be raised with regard to the naturalness of this procedure. In this section, we will attempt to address such objections, as well as to make contact with other possible evaluations of the action.

In the literature, when (non-null) bounded regions are considered, the boundary's flip surfaces are often hidden in topological corners. Examples include purely-spacelike and purely-timelike boundaries, as well as mixed-signature ``cylinders'' with spacelike bases and timelike sides, as in figure \ref{fig:mixed_boundary}. For such boundaries, it is tempting to simply add up the contributions from the spacelike or timelike smooth patches, without ever including the corner contributions. However, this recipe is not valid for two reasons. First, it was shown \cite{Hayward:1993my} that the corner contributions are required for a variational principle where one keeps only the boundary metric fixed. Second, the recipe does not extend well to smooth boundaries, as we will now discuss.

Consider a smooth boundary in 1+1d (figure \ref{fig:angles_smooth_boundary}). By the ``corner ignoring'' recipe, one would want to add up the boundary contributions from the spacelike and timelike patches, without integrating through the signature flips. Now, since the boost angle diverges at the flips, these separate contributions are all divergent. We are thus \emph{forced} to carry out the boundary integral through the flips, in order to regularize these divergences. The result is that the real infinities from the two sides of each flip cancel, at the cost of picking up an imaginary part.

Now, in \cite{Hayward:1993my}, after asserting the necessity of corner terms, a recipe for the action is given which is precisely \emph{the real part} of the action as understood in the present paper. In the language of section \ref{sec:GR:contour}, this corresponds to taking the average of the two complex-conjugate contour integrals, instead of choosing one of them. When phrased in this way, it seems like a suspect procedure. Indeed, quantum mechanics tells us that one should sum (or average) over the amplitudes $e^{iS}$ for different configurations, but not over the actions $S$ themselves.  

In fact, the real action in \cite{Hayward:1993my} is obtained not by any sort of integral through the flip surfaces (though the integral \emph{is} evaluated for corners that don't involve signature flips). Instead, it is arrived at by demanding a variational principle with fixed boundary metric. It is important to note that our imaginary part of the action has \emph{no effect} on this variational principle. Indeed, $\Im S$ depends only on the areas of the flip surfaces, which are a property of the boundary metric. Unlike the real parts of corner angles, the imaginary boost angles $\pi i/2$ are constant, and one doesn't have to worry about their variation.

Similarly, there is no contradiction between our complex action and a real Hamiltonian. First, one must realize that for finite regions in Lorentzian spacetime, the relation between the action and a Hamiltonian formalism is somewhat subtle: we must first have a suitable notion of ``time evolution'' that would respect causality. The simplest way to arrange this is to consider as ``time slices'' a family of spacelike hypersurfaces with a shared codimension-2 boundary $\Sigma$. Under ``time evolution'', the hypersurfaces deform into each other, leaving the intrinsic geometry of $\Sigma$ fixed. Any two hypersurfaces form a closed spacelike boundary, as in figure \ref{fig:spacelike_boundary}. The signature flips are hidden in the corner $\Sigma$, and the action's imaginary part is given in terms of its area by \eqref{eq:ImS_spacelike}. Thus, under this sort of time evolution, $\Im S$ is constant, i.e. the action's variation is \emph{real}. This is consistent with the usual notion of a real Hamiltonian.  

\section{The imaginary part of the Lovelock action} \label{sec:Lovelock}

In this section, we extend our analysis to Lovelock theories of gravity \cite{Lovelock:1971yv}. The action takes the form:
\begin{align}
 S = \sum_{m=0}^{\lfloor d/2 \rfloor} c_m S_m \ . \label{eq:Lovelock_S}
\end{align}
The ``order'' $m$ of the action terms is bounded by half the spacetime dimension. The $c_m$ are constant coefficients. The $S_m$ are geometric quantities of dimension $(length)^{d-2m}$ that consist of a bulk integral plus a boundary integral:
\begin{align}
 S_m = \int_\Omega \calL_m\, d^dx + \int_{\del\Omega} Q_m\, d^{d-1}x \ . \label{eq:Lovelock_S_m}
\end{align}
The $m$'th-order Lagrangian $\calL_m$ is given by:
\begin{align}
 \calL_m = \frac{(2m)!}{2^m}\sqrt{-g}\,R^{[\mu_1\nu_1}{}_{[\mu_1\nu_1}R^{\mu_2\nu_2}{}_{\mu_2\nu_2}\dots R^{\mu_m\nu_m]}{}_{\mu_m\nu_m]} \ , 
 \label{eq:Lovelock_L}
\end{align}
where the upper and lower indices are antisymmetrized before tracing (actually, it's sufficient to antisymmetrize one of the two sets). The numerical coefficient is chosen to conform with the conventions of \cite{Myers:1987yn,Jacobson:1993xs}. It arises in the transition from Cartan notation. The $(2m)!$ factor counts the permutations of the $2m$ antisymmetrized indices. The $1/2^m$ factor corrects for the permutations within each index pair $(\mu_k \nu_k)$.

As an example, the three lowest-order Lagrangians read:
\begin{align}
 \begin{split}
   \calL_0 &= \sqrt{-g} \\
   \calL_1 &= \sqrt{-g}R \\ 
   \calL_2 &= \sqrt{-g}(R^2 - 4R_{\mu\nu}R^{\mu\nu} + R_{\mu\nu\rho\sigma}R^{\mu\nu\rho\sigma}) \ .
 \end{split}
\end{align}
Standard GR with a cosmological constant is obtained by setting $c_0 = \Lambda/(16\pi G)$ and $c_1 = 1/(16\pi G)$, with all the higher $c_m$ vanishing.

\subsection{Rewriting the boundary term}

As in GR, the imaginary contribution to the action will come from extrinsic curvature factors in the boundary integrand $Q_m$. The expression for $Q_m$ was found by Myers in \cite{Myers:1987yn}, where it is given in Cartan notation. For our purposes, Riemannian notation is more convenient. It will allow us to keep the upper and lower indices in \eqref{eq:Lovelock_L} on an equal footing, and will also simplify the dependence on the extrinsic curvature. For now, we consider non-null patches of the boundary. 

In \cite{Myers:1987yn}, the extrinsic curvature is given in terms of the second fundamental form $\theta^A{}_B = \theta^A{}_{BC}E^C$, where $(A,B,C)$ are internal orthonormal indices, and $E^C$ is the vielbein 1-form. Converting all internal indices into world-indices, we write this as $\theta^\mu{}_{\nu\rho}$. We prefer to work in terms of the Riemannian extrinsic curvature $K_\mu^\nu$, defined as the boundary projection of $\nabla_\mu n^\nu$, where $n_\mu$ is an outgoing normal covector. The two tensors are related as follows: 
\begin{align}
 K_\mu^\nu &= \theta^\mu{}_{\rho\nu}n^\rho \\
 \theta^{\mu\nu}{}_\rho &= \frac{2}{n\cdot n}K_\rho^{[\mu} n^{\nu]} \ . \label{eq:theta}
\end{align}
The $n\cdot n$ factor in the denominator makes \eqref{eq:theta} valid for normals $n^\mu$ of arbitrary norm and signature. This will facilitate the study of the null limit.

The next ingredient of the formula for $Q_m$ is the curvature form $\Omega_s$ (which we will write as a tensor $R_s$) of an auxiliary connection $\omega_s = \omega_0 + s\theta$. As $s$ runs from 0 to 1, this interpolates between the boundary's intrinsic connection $\omega_0$ and the bulk connection $\omega_1 = \omega_0 + \theta$. We will only need the components of $R_s$ tangent to the boundary. These are given by:
\begin{align}
 (R_s)^{ab}{}_{cd} = (R_0)^{ab}{}_{cd} + 2s^2\theta^a{}_{\mu [c}\theta^{\mu b}{}_{d]} 
   = (R_0)^{ab}{}_{cd} - \frac{2s^2}{n\cdot n} K^{[a}_{[c} K^{b]}_{d]} \ , \label{eq:R_s_raw}
\end{align}
where $(R_0)^{ab}{}_{cd}$ is the boundary's intrinsic curvature. In the null limit, the intrinsic connection and curvature become ill-defined. Therefore, we prefer to express everything in terms of the \emph{bulk} Riemann curvature $R^{ab}{}_{cd} = (R_1)^{ab}{}_{cd}$ and the extrinsic curvature $K_a^b$. Thus, we rewrite \eqref{eq:R_s_raw} as:
\begin{align}
 (R_s)^{ab}{}_{cd} = R^{ab}{}_{cd} + \frac{2(1 - s^2)}{n\cdot n} K^{[a}_{[c} K^{b]}_{d]} \ . \label{eq:R_s}
\end{align}

After these preliminaries, the result of \cite{Myers:1987yn} for the boundary integrand $Q_m$ reads:
\begin{align}
 Q_m = \frac{(2m)!}{2^{m-1}}\sqrt{\frac{-h}{n\cdot n}}\int_0^1 ds\, 
   K_{[a}^{[a} (R_s)^{b_1 c_1}{}_{b_1 c_1} (R_s)^{b_2 c_2}{}_{b_2 c_2}\dots (R_s)^{b_{m-1} c_{m-1}]}{}_{b_{m-1} c_{m-1}]} \ . \label{eq:Q_raw}
\end{align}
The numerical coefficient $(2m)!/2^{m-1}$ arises as follows: a factor of $m$ comes from the result as written in \cite{Myers:1987yn}; a factor of 2 from the two possible placements of the normal index in \eqref{eq:theta}; a factor of $(2m-1)!$ from the permutations of the $2m-1$ antisymmetrized boundary indices; and a factor of $1/2^{m-1}$ to correct for the permutations within each index pair $(b_k c_k)$. The density factor $\sqrt{-h/(n\cdot n)}$ and the overall sign are the same as in \eqref{eq:GR_action}. They arise from the boundary volume density and the translation \eqref{eq:theta} between $\theta^{\mu\nu}{}_\rho$ and $K_a^b$.

Finally, let us substitute the expression \eqref{eq:R_s} for $R_s$ into \eqref{eq:Q_raw} and perform the auxiliary $s$ integral. Expanding the product of $R_s$ tensors is straightforward, since the antisymmetry in both upper and lower indices makes all $R^{ab}{}_{cd}$ factors and all $K_a^b$ factors interchangeable. The expression for $Q_m$ becomes:
\begin{align}
 \begin{split}
   Q_m = \frac{(2m)!}{2^{m-1}}&\sqrt{\frac{-h}{n\cdot n}}\,\sum_{p=0}^{m-1} \binom{m-1}{p} \frac{2^{3p} (p!)^2}{(2p+1)!(n\cdot n)^p} \\
     &\times K_{[a_1}^{[a_1} K_{a_2}^{a_2}\dots K_{a_{2p+1}}^{a_{2p+1}} 
     R^{b_1 c_1}{}_{b_1 c_1} R^{b_2 c_2}{}_{b_2 c_2}\dots R^{b_{m-p-1} c_{m-p-1}]}{}_{b_{m-p-1} c_{m-p-1}]} \ ,
 \end{split} \label{eq:Q} 
\end{align}
where we used the binomial expansion and the integral formula:
\begin{align}
 \int_0^1 (1 - s^2)^p ds = \frac{2^{2p} (p!)^2}{(2p+1)!} \ . 
\end{align}

As an example, at the three lowest orders, eq. \eqref{eq:Q} evaluates to:
\begin{align}
 \begin{split}
   Q_0 &= 0 \\
   Q_1 &= 2\sqrt{\frac{-h}{n\cdot n}}\,K \\
   Q_2 &= 4\sqrt{\frac{-h}{n\cdot n}}\left(KR - 2K_{ab}R^{ab} + \frac{2}{3n\cdot n}(K^3 - 3KK_{ab}K^{ab} + 2K_a^b K_b^c K_c^a) \right) \ . 
 \end{split}
\end{align}
In \cite{Myers:1987yn}, these expressions were given with a slight error: in the expression for $Q_2$, the factor of 3 in front of $KK_{ab}K^{ab}$ is missing.

\subsection{Corner terms}

Retracing the argument in section \ref{sec:GR}, let us now find the formula for corner contributions to the Lovelock action. Suppose that the boundary bends sharply by an angle $\alpha$ along some corner surface $\Sigma$. This will produce a singularity in the component $K_\bot^\bot$ of $K_a^b$ ``along the bend''. The contribution of this singularity to the integral of $Q_m$ over the boundary is a simple generalization of eq. \eqref{eq:K_corner}. To write it down, let us isolate a $K_\bot^\bot$ factor from $Q_m$ as follows:
\begin{align}
 Q_m \equiv \sqrt{\frac{-h}{n\cdot n}}\,K_\bot^\bot\, I_m + (\text{terms with no factors of } K_\bot^\bot) \ . \label{eq:isolate}
\end{align}
The corner contribution, i.e. the integral of $Q_m$ over a small neighborhood $\delta\Sigma$ of the corner, then reads:
\begin{align}
 \int_{\delta\Sigma}Q_m\, d^{d-1}x = \int_\Sigma \sqrt{\gamma}\, d^{d-2}x\int d\alpha\, I_m \ . \label{eq:Q_corner_raw}
\end{align}
Here, $\gamma$ is the determinant of the corner's codimension-2 metric $\gamma_{ij}$, and $\alpha$ is the angle along the bend. The $d\alpha$ integral cannot be carried out immediately, for a reason that will soon become apparent.   

Let us evaluate the quantity $I_m$ in \eqref{eq:isolate}. Due to the antisymmetrization in \eqref{eq:Q}, each term in $Q_m$ contains at most one factor of $K_\bot^\bot$. All other indices must be tangent to the corner surface. We denote such indices by $(i,j,\dots)$. The component $K_\bot^\bot$ can be any of the $2p+1$ factors of $K_a^b$ in \eqref{eq:Q}. This yields a combinatoric factor of $2p+1$. Once we extract $K_\bot^\bot$ from the antisymmetrized product in \eqref{eq:Q}, the number of antisymmetrized indices is reduced from $2m-1$ to $2m-2$. The ratio of the two implicit factorials yields another combinatoric factor of $1/(2m-1)$. With this in mind, $I_m$ can be read off from \eqref{eq:Q} as:
\begin{align}
 \begin{split}
   I_m = {}& 2m\frac{(2m-2)!}{2^{m-1}}\sum_{p=0}^{m-1} \binom{m-1}{p} \frac{2^{3p} (p!)^2}{(2p)!(n\cdot n)^p} \\
     &\times K_{[i_1}^{[i_1} K_{i_2}^{i_2}\dots K_{i_{2p}}^{i_{2p}} 
     R^{j_1 k_1}{}_{j_1 k_1} R^{j_2 k_2}{}_{j_2 k_2}\dots R^{j_{m-p-1} k_{m-p-1}]}{}_{j_{m-p-1} k_{m-p-1}]} \ ,
 \end{split}
\end{align}
and the corner contribution \eqref{eq:Q_corner_raw} becomes:
\begin{align}
 \begin{split}
   \int_{\delta\Sigma}Q_m\, d^{d-1}x ={}& 2m\frac{(2m-2)!}{2^{m-1}}
     \int_\Sigma \sqrt{\gamma}\, d^{d-2}x \int d\alpha \sum_{p=0}^{m-1} \binom{m-1}{p} \frac{2^{3p} (p!)^2}{(2p)!(n\cdot n)^p} \\
     &\times K_{[i_1}^{[i_1} K_{i_2}^{i_2}\dots K_{i_{2p}}^{i_{2p}} 
     R^{j_1 k_1}{}_{j_1 k_1} R^{j_2 k_2}{}_{j_2 k_2}\dots R^{j_{m-p-1} k_{m-p-1}]}{}_{j_{m-p-1} k_{m-p-1}]} \ . \label{eq:Q_corner}
 \end{split}  
\end{align}
Naively, the extrinsic curvature factors $K_i^j$ in \eqref{eq:Q_corner} are ill-defined, since they may be discontinuous at the corner. However, it is possible to give meaning to $K_i^j$ as a function of the angle $\alpha$ along the bend. This is why we kept track of the $d\alpha$ integral.

Let us specialize to a spacelike corner $\Sigma$, which is the relevant case for signature flips. At each point of $\Sigma$, the 1+1d plane orthogonal to it is spanned by two null vectors $L^\mu$ and $\ell^\mu$, each with an arbitrary scaling. $L^\mu$ and $\ell^\mu$ can be viewed as null normals to the two unique lightsheets containing $\Sigma$. As we traverse the corner ``along the bend'', the boundary normal $n^\mu$ gets rotated (and rescaled) in the plane spanned by $L^\mu,\ell^\mu$. We can thus write $n^\mu$ in the form \eqref{eq:n_L_ell}:
\begin{align}
 n^\mu = \lambda(L^\mu + z\ell^\mu); \quad n\cdot n = 2\lambda^2 z (L\cdot\ell) \ . \label{eq:n_L_ell_Lovelock}
\end{align}
We imagine $z$ changing continuously from an initial value $z_1$ on one side of the corner to a final value $z_2$ on the other side. $\lambda$ may change as well, but this will have no effect on the calculation. As in \eqref{eq:d_alpha}, the change in $z$ is related to the normal's boost angle as $d\alpha = \pm dz/2z$, where the sign depends on orientation conventions. In particular, the endpoint ratio $z_2/z_1$ is related to the overall corner angle $\alpha$ as $z_2/z_1 = e^{\pm 2\alpha}$. The overall scaling of $z_1$ and $z_2$ depends on the arbitrary scalings of $L^\mu$ and $\ell^\mu$.

Now, the extrinsic curvature components $K_i^j$ in \eqref{eq:Q_corner} can be written as:
\begin{align}
 K_i^j = \nabla_i n^j = \nabla_i(\lambda(L^j + z\ell^j)) = \lambda(\nabla_i L^j + z\nabla_i\ell^j) \ , \label{eq:K_ij}
\end{align}
where we used the fact that $L^i = \ell^i = 0$, since $L^\mu$ and $\ell^\mu$ are orthogonal to $\Sigma$. The derivatives $\nabla_i L^j$ and $\nabla_i\ell^j$ are the shear/expansion tensors of the two lightsheets generated by $L^\mu$ and $\ell^\mu$. Using eqs. \eqref{eq:n_L_ell_Lovelock}-\eqref{eq:K_ij}, the $z$-dependent factors in \eqref{eq:Q_corner} become:
\begin{align}
 \begin{split}
   \int\frac{d\alpha}{(n\cdot n)^p} K_{[i_1}^{[i_1}\dots K_{i_{2p]}}^{i_{2p]}}
     = \pm\int_{z_1}^{z_2}&\frac{dz}{(2z)^{p+1}(L\cdot \ell)^p}\sum_{r=0}^{2p} \binom{2p}{r} z^r \\ 
     &\times \nabla_{[i_1}L^{[i_1}\dots \nabla_{i_{2p-r}}L^{i_{2p-r}} \nabla_{j_1}\ell^{j_1}\dots \nabla_{j_r]}\ell^{j_r]} \ .
 \end{split} \label{eq:z_parts}
\end{align}
The $dz$ integral can now be performed, and the result related back to the boundary normals $n^\mu_{1,2}\sim L^\mu + z_{1,2}\ell^\mu$ before and after the corner. Plugging back into \eqref{eq:Q_corner}, we will get the corner contribution to the action. A similar analysis can be performed for timelike corners, using a pair of \emph{complex} null normals. Rather than following through these steps, we now turn to our true object of interest - the imaginary contribution to the action at a flip surface. 

\subsection{Contributions to $\Im S$ from corners with signature flips}

We now wish to apply eqs. \eqref{eq:Q_corner} and \eqref{eq:z_parts} to signature-flip surfaces, where the bending angle has an imaginary part $\Im\alpha = \pi/2$. For GR, we started off by considering smooth boundaries in section \ref{sec:GR:smooth}, and treated boundaries with topological corners as a limiting case. For higher-order Lovelock gravity, it turns out that boundaries with corners are the easier case. Thus, we will first consider a mixed-signature boundary with corners at the flip surfaces, as in figure \ref{fig:mixed_boundary}. We will come back to smooth boundaries in section \ref{sec:Lovelock:smooth}.

Let $\Sigma$ be a corner surface containing a signature flip, and let $L^\mu,\ell^\mu$ be the two null vectors orthogonal to it. As in section \ref{sec:GR:contour}, the signature flip means that $z$ in the angular integral \eqref{eq:z_parts} crosses through zero. Thus, the endpoints $z_{1,2}$ in \eqref{eq:z_parts} should be taken as real numbers, one negative and the other positive. Again as in section \ref{sec:GR:contour}, we avoid the singularity at $z=0$ by deforming the integration contour into the complex plane. As a result, the integral \eqref{eq:z_parts} picks up an imaginary part, given by a closed contour integral divided by $2i$:
\begin{align}
 \begin{split}
   \Im\int\frac{d\alpha}{(n\cdot n)^p} K_{[i_1}^{[i_1}\dots K_{i_{2p]}}^{i_{2p]}}
       = \pm\frac{1}{2i}\oint&\frac{dz}{(2z)^{p+1}(L\cdot \ell)^p}\sum_{r=0}^{2p} \binom{2p}{r} z^r \\ 
       &\times \nabla_{[i_1}L^{[i_1}\dots \nabla_{i_{2p-r}}L^{i_{2p-r}} \nabla_{j_1}\ell^{j_1}\dots \nabla_{j_r]}\ell^{j_r]} \ .
 \end{split}
\end{align}
The integral selects the $r=p$ term, and we get: 
\begin{align}
  \Im\int\frac{d\alpha}{(n\cdot n)^p} K_{[i_1}^{[i_1}\dots K_{i_{2p]}}^{i_{2p]}} = \pm\frac{\pi}{2^{p+1}(L\cdot \ell)^p} \binom{2p}{p}
    \nabla_{[i_1}L^{[i_1}\dots \nabla_{i_p}L^{i_p} \nabla_{j_1}\ell^{j_1}\dots \nabla_{j_p]}\ell^{j_p]} \ . \label{eq:Im_int}
\end{align}
Plugging into \eqref{eq:Q_corner}, the imaginary action contribution from the flip surface reads:
\begin{align}
 \begin{split}
   \Im\int_{\delta\Sigma}&Q_m\, d^{d-1}x = \pm \pi m\frac{(2m-2)!}{2^{m-1}}
     \int d^{d-2}x\,\sqrt{\gamma}\sum_{p=0}^{m-1} \binom{m-1}{p} \frac{2^{2p}}{(L\cdot\ell)^p} \\
     &\times \nabla_{[i_1}L^{[i_1}\dots \nabla_{i_p}L^{i_p} \nabla_{j_1}\ell^{j_1}\dots \nabla_{j_p}\ell^{j_p} 
     R^{k_1 l_1}{}_{k_1 l_1}\dots R^{k_{m-p-1} l_{m-p-1}]}{}_{k_{m-p-1} l_{m-p-1}]} \ .
 \end{split} \label{eq:Im_Q_raw}  
\end{align}
The $(2p)!$ and $p!$ factors from \eqref{eq:Q_corner} got canceled by those in \eqref{eq:Im_int}. As a result, the sum over $p$ in \eqref{eq:Im_Q_raw} has the form of a binomial expansion. The expression thus simplifies into: 
\begin{align}
 \begin{split}
   \Im\int_{\delta\Sigma}Q_m\, d^{d-1}x = \pm &\pi m\frac{(2m-2)!}{2^{m-1}} \int_\Sigma d^{d-2}x\,\sqrt{\gamma} \\
     &\times \tilde R^{[i_1 j_1}{}_{[i_1 j_1}\tilde R^{i_2 j_2}{}_{i_2 j_2}\dots \tilde R^{i_{m-1} j_{m-1}]}{}_{i_{m-1} j_{m-1}]} \ ,
 \end{split} \label{eq:Im_Q_medium}  
\end{align}
where:
\begin{align}
 \tilde R^{ij}{}_{kl} = R^{ij}{}_{kl} + \frac{4}{L\cdot\ell}\nabla_{[k} L^{[i}\nabla_{l]}\ell^{j]} \ . \label{eq:tilde_R}
\end{align}
This combination is the just intrinsic codimension-2 Riemann tensor of the flip surface $\Sigma$. Thus, eq. \eqref{eq:Im_Q_medium} can be written as:
\begin{align}
 \Im\int_{\delta\Sigma}Q_m\, d^{d-1}x = \pm &\pi m \int_\Sigma \tilde\calL_{m-1}\, d^{d-2}x \ ,
\end{align}
where $\tilde\calL_m$ is the codimension-2 Lovelock density analogous to \eqref{eq:Lovelock_L}, with $\sqrt{\gamma}$ and $\tilde R^{ij}{}_{kl}$ in place of $\sqrt{-g}$ and $R^{\mu\nu}{}_{\rho\sigma}$.

Plugging into eqs. \eqref{eq:Lovelock_S}-\eqref{eq:Lovelock_S_m}, we obtain the imaginary part of the Lovelock action:
\begin{align}
 \Im S = \pm\pi\sum_{\mathrm{flips}}\sum_{m=1}^{\lfloor d/2 \rfloor} c_m \int\tilde\calL_{m-1}\, d^{d-2}x \ . \label{eq:ImS_Lovelock}
\end{align}
As in section \ref{sec:GR:contour}, the overall sign in \eqref{eq:ImS_Lovelock} is ultimately controlled by whether we bypass the pole in \eqref{eq:z_parts} from above or from below. The choice should be made such that $\Im S \geq 0$, so that quantum amplitudes $e^{iS}$ don't explode exponentially. Assuming that the GR ($m=1$) term dictates the sign of the entire sum \eqref{eq:ImS_Lovelock}, and assuming that $c_1 = 1/(16\pi G)$ is positive, the sign in \eqref{eq:ImS_Lovelock} should be plus. The $m=1$ term then reproduces the GR result \eqref{eq:ImS_smooth}. Moreover, the sum \eqref{eq:ImS_Lovelock} then satisfies the relation \eqref{eq:result} with the entropy formula \cite{Jacobson:1993xs} for Lovelock gravity. 

In particular, we see that the assumption that a plus sign should be chosen in \eqref{eq:ImS_Lovelock} is equivalent to the positivity of the entropy formula. This is guaranteed to hold if the GR term in \eqref{eq:ImS_Lovelock} is much larger than the rest, which is always the case at sufficiently large distance scales. We stress that even with this stronger assumption, our calculation is non-stationary. The assumption implies that gradients are small with respect to the coupling ratios $c_1/c_{m>1}$, while stationarity means that time gradients are small with respect to space gradients.

Finally, we note that the imaginary part \eqref{eq:ImS_Lovelock} depends only on the intrinsic metric of the flip surfaces. This means that as in GR, $\Im S$ has no effect on the variational principle with fixed boundary metric (section \ref{sec:compare}). 

\subsection{Smooth boundaries as a limiting case} \label{sec:Lovelock:smooth}

For flip surfaces $\Sigma$ not associated with a corner, as in section \ref{sec:GR:smooth}, a problem arises. In Lovelock theories with $m>1$, the boundary term \eqref{eq:Q} can also give divergent contributions from terms that \emph{don't} include a factor of $K_\bot^\bot$. These divergences arise from the negative powers $(n\cdot n)^{-p}$ of the normal's length in the $p>0$ terms. Upon deforming the integration contour, these terms may lead to additional imaginary contributions to the action. 

Such additional contributions to $\Im S$ are undesirable for two reasons. First, they would spoil the beautiful result \eqref{eq:ImS_Lovelock}, which doesn't depend on the shear/expansion tensors $\nabla_i L^j$ and $\nabla_i\ell^j$ of the two lightsheets orthogonal to $\Sigma$ (we note, however, that the result still holds if the shear/expansion tensor $\nabla_i L^j$ vanishes on $\Sigma$). Second, unlike the contributions with a factor of $K_\bot^\bot$, they cannot be written universally as an integral over the boost angle. As a result, they would depend on the boundary's detailed shape in the neighborhood of $\Sigma$. 

We propose that the solution to these problems is to always treat a flip surface as the limiting case of a corner. As discussed in section \ref{sec:GR:spacelike_timelike}, this is possible because a fixed angle between a timelike and a spacelike vector (with imaginary part $\pi/2$) can appear arbitrarily blunt, as the two vectors are boosted towards a null direction. In other words, by an infinitesimal deformation of the boundary, one can always turn a smooth signature flip into a corner. When this is done, the potentially divergent boundary-term contributions with no factors of $K_\bot^\bot$ vanish, due to the order of limits: although $(n\cdot n)^{-p}$ tends to infinity at the flip surface ``somewhere inside'' the corner, this divergence is always multiplied by the corner's strictly vanishing width.

With this understanding, the result \eqref{eq:ImS_Lovelock} holds also for smooth boundaries. We note the subtle yet consistent use of limiting procedures involved. To evaluate the corner contributions from terms in $Q_m$ with factors of $K_\bot^\bot$, we treat corners as a limiting case of smooth curves. On the other hand, to eliminate extra imaginary contributions from terms in $Q_m$ with \emph{no} factors of $K_\bot^\bot$, we treat smooth signature flips as a limiting case of corners. 

\subsection{Boundaries of fixed causal type and stationary black holes} \label{sec:Lovelock:black_hole}

The formula \eqref{eq:ImS_Lovelock} extends straightforwardly to corners containing more than one signature flip. As in section \ref{sec:GR:spacelike_timelike}, each corner contributes to $\Im S$ according to the number of signature flips inside it, i.e. that would appear if one resolved the corner into a smooth curve. For the purely-spacelike and purely-timelike boundaries from section \ref{sec:GR:spacelike_timelike}, each corner contains two flips. Thus, for such boundaries, eq. \eqref{eq:ImS_Lovelock} reads: 
\begin{align}
 \Im S = 2\pi\sum_{\mathrm{corners}}\sum_{m=1}^{\lfloor d/2 \rfloor} c_m \int\tilde\calL_{m-1}\, d^{d-2}x \ , \label{eq:ImS_Lovelock_corners}
\end{align}
where we assume that the plus sign is chosen in \eqref{eq:ImS_Lovelock}, i.e. that \eqref{eq:ImS_Lovelock_corners} is positive.

For purely-null boundaries, as in section \ref{sec:GR:null}, the key difficulty is to fix the locations of the flip surfaces in \eqref{eq:ImS_Lovelock}. We again imagine a path integral over the different possibilities. This picks out the choice that minimizes $\Im S$. While in GR we had a clean theorem about the area's monotonicity, no such theorem is known for the functional \eqref{eq:ImS_Lovelock} in Lovelock gravity. To retain the GR conclusion that the flip surfaces should be placed at the tips, we must assume that the monotonicity of the sum \eqref{eq:ImS_Lovelock} as one travels along the boundary lightsheets is dictated by the $m=1$ term. As with the choice of sign in \eqref{eq:ImS_Lovelock}, this is guaranteed at sufficiently large distance scales, where the GR term dominates over the $m>1$ terms. In that case, $\Im S$ is given by a variant of \eqref{eq:ImS_Lovelock_corners}:
\begin{align}
 \Im S = 2\pi\sum_{\mathrm{tips}}\sum_{m=1}^{\lfloor d/2 \rfloor} c_m \int\tilde\calL_{m-1}\, d^{d-2}x \ , \label{eq:ImS_Lovelock_null}
\end{align}
with the factor of 2 arising from the two flips near each tip point. For lightsheets with a stationary intrinsic metric, i.e. zero shear and expansion, the remarks at the end of section \ref{sec:GR:null} apply. In particular, since the integral in \eqref{eq:ImS_Lovelock} depends only on the flip surface's intrinsic metric, the location of the flip surface along the lightsheet doesn't affect $\Im S$. 

A direct connection with the entropy of stationary black holes can be made as in section \ref{sec:GR:black_hole}. With our assumptions on the sign and (for null boundaries) the monotonicity of $\Im S$, we can apply eqs. \eqref{eq:ImS_Lovelock_corners}-\eqref{eq:ImS_Lovelock_null} to the special boundaries in figure \ref{fig:kruskal}. The sum over corners again yields a factor of 2. The result reads:
\begin{align}
 \Im S = 4\pi\sum_{m=1}^{\lfloor d/2 \rfloor} c_m \int\tilde\calL_{m-1}\, d^{d-2}x \ , \label{eq:Lovelock_entropy}
\end{align}
where the integral is over any horizon slice. This reproduces the entropy formula from \cite{Jacobson:1993xs}.

\section{Discussion} \label{sec:discuss}

In this paper, we studied the imaginary part of the gravitational action in finite regions. This arose from the action's boundary term, with a crucial role played by signature flips. The imaginary action for arbitrary regions is closely related to the entropy formula, as in eq. \eqref{eq:result}. In the special case of stationary spacetimes, we identified a class of regions for which the imaginary action directly reproduces the black hole entropy. These results draw a new link between causality, black hole thermodynamics and the dynamics of gravity in finite regions. Our emphasis on finite regions and non-stationarity may prove fruitful, since these two features are central in an accelerated-expansion cosmology.

We performed the calculation for GR with minimally-coupled matter, as well as for Lovelock gravity. The outcome of the Lovelock calculation gives us confidence in the validity of the methods used. In particular, it's encouraging how the simple result \eqref{eq:ImS_Lovelock}, depending only on the intrinsic geometry of the flip surface, came about through some intricate cancellations. This raises the question of whether the result can be derived in a simpler manner. In particular, one would want to take advantage of the topological structure of the action terms, emphasized in \cite{Myers:1987yn}. It seems that this is indeed possible, by drawing a relation with the Euclidean result of \cite{Banados:1993qp} (I am grateful to the anonymous referee at JHEP for pointing this out). The detailed argument will be presented in future work, together with other interrelations between the Lorentzian and Euclidean calculations.

The general result \eqref{eq:ImS_Lovelock} also lends credence to \eqref{eq:Lovelock_entropy} as the correct non-stationary entropy formula for Lovelock gravity. This agrees with the proposal \cite{Iyer:1994ys} for a non-stationary entropy in general gravitational theories. Our result provides further motivation to look for an increase law for this quantity - an issue that we avoided by assuming that the GR term dominates. In this context, see the partial result \cite{Kolekar:2012tq}. On the other hand, see the argument in \cite{Sarkar:2010xp} about a possible decrease of the quantity \eqref{eq:Lovelock_entropy} in black hole mergers.  

An explicit argument linking our imaginary actions to entropy is still missing. If found, it would upgrade our non-stationary calculation of $\Im S$ into a non-stationary derivation of gravitational entropy. On a related note, it's not clear how to generalize our treatment of black holes (sections \ref{sec:GR:black_hole},\ref{sec:Lovelock:black_hole}) outside the stationary case. For realistic, non-stationary black holes, there is no bifurcate horizon. No analogues of the timelike or null boundaries from figure \ref{fig:kruskal} are apparent. One \emph{can} draw an analogue of the spacelike boundary from figure \ref{fig:kruskal}(a). However, such a boundary would have just one horizon slice as its corner, giving $\Im S = A/8G$ instead of $A/4G$. On the whole, the scheme presented in sections \ref{sec:GR:black_hole},\ref{sec:Lovelock:black_hole} is too non-local for non-stationary black holes. A scheme more adapted to such spacetimes appears to exist, and will be reported separately.  

\subsection{$\Im S$ vs. entropy}

One possibility is that after all, there is no general concept of gravitational entropy. Instead, there is just the imaginary part $\Im S$ of the action. We know that this quantity obeys equations analogous to those for entropy, i.e. black hole thermodynamics. We are also quite certain that it describes an actual entropy in special cases - stationary black holes and AdS/CFT \cite{Maldacena:1997re,Aharony:1999ti}. However, it is possible that that's all there is to it. We will now provide two speculative arguments in favor of this view.

Recall that through Wald's construction \cite{Wald:1993nt}, entropy is related to the flux of the Noether potential of a diffeomorphism. Consider a simpler quantity of this sort - the Noether potential $\sqrt{-g}F^{\mu\nu}$ of the electromagnetic gauge group in Einstein-Maxwell theory. By the Gauss law, the flux of this quantity through a codimension-2 surface measures the charge inside. But does its flux through a small surface element measure a charge \emph{density}? Moreover, does its flux through a \emph{hyper}surface element measure a current density? In general, the answer is negative, but not always. For instance, a stationary black hole horizon viewed from the outside is indistinguishable from a charged membrane, with charge and current densities given by the local flux of $\sqrt{-g}F^{\mu\nu}$. Moreover, in a certain quasi-stationary limit, these densities satisfy a conservation law and experience a Lorentz force. This is part of the ``membrane paradigm'' of black hole hydrodynamics \cite{Damour,Price:1986yy,Eling:2010hu}.

$\sqrt{-g}F^{\mu\nu}$ also describes a physical charge/current density when one of the system's dimensions is much shorter than any wavelengths in the other dimensions. For example, consider a plate capacitor with spacing $D$ between the plates and a charge density that varies over distances $\lambda\gg D$. Then the flux of $\sqrt{-g}F^{\mu\nu}$ in perpendicular to the plates, i.e. the electric field, corresponds directly to the local charge density. Another example is AdS/CFT. There, the role of ``capacitor plates'' is played by the AdS boundary, and the radial distance, though infinite, diverges much slower than any tangential wavelengths. As a result, the flux of $\sqrt{-g}F^{\mu\nu}$ at the boundary corresponds to a current density in the CFT.

To sum up, the Noether potential $\sqrt{-g}F^{\mu\nu}$ obeys the same equations as a current density on a hypersurface, and arguably \emph{is} a current density for stationary black holes and in AdS/CFT. However, in general, Einstein-Maxwell theory does not have a concept of current on hypersurfaces. Instead, there is just the Noether potential. We propose that a similar situation may hold for the imaginary part of the action vs. gravitational entropy: the imaginary action is the general concept, which corresponds to an entropy only in special cases.

From another perspective, if one treats holography with full seriousness, the foundations of quantum mechanics become suspect. Since finite regions can hold only a finite amount of information, one cannot measure e.g. quantum probabilities with arbitrary precision. Black holes, as well as positive-$\Lambda$ cosmology, are situations where we are confined to such a finite region. Thus, the notions of Hilbert space, unitarity, etc. may lose their sharp meanings. If this is the case, then the same can be said about entropy. In this view, the concept of gravitational entropy is much like Bohr's electron orbits: it contains an important grain of truth, but it's formulated in a language which this very truth makes obsolete. The imaginary action may represent a step towards more appropriate concepts.     

\section*{Acknowledgements}		

I am grateful to Abhay Ashtekar and Norbert Bodendorfer for discussions, and to Steve Carlip for an email exchange. This work is supported in part by the NSF grant PHY-1205388 and the Eberly Research Funds of Penn State.

\end{document}